# Network Functional Compression for Control Applications

Sifat Rezwan*, Juan A. Cabrera*, and Frank H. P. Fitzek*†
*Deutsche Telekom Chair of Communication Networks, TU Dresden, Germany
†Centre for Tactile Internet with Human-in-the-Loop (CeTI)
{sifat.rezwan | juan.cabrera | frank.fitzek}@tu-dresden.de

*Abstract*—The trend of future communication systems is to aim for the steering and control of cyber-physical systems. These systems can quickly become congested in environments like those presented in Industry 4.0. In these scenarios, a plethora of sensor data is transmitted wirelessly to multiple in-network controllers that compute the control functions of the cyber-physical systems. In this paper, we show an implementation of network Functional Compression (FC) as a proof of concept to drastically reduce the data traffic in these scenarios. FC is a form of goal-oriented communication scheme in which the objective of the sender-receiver pair is to transmit the minimum amount of information to compute a function at the receiver end. In our scenario, the senders transmit an encoded and compressed version of the sensor data to a destination, an in-network controller interested in computing as its target function, a PID controller. We show that it is possible to achieve compression rates of over 50% in some cases by employing FC. We also show that using FC in a distributed cascade fashion can achieve more significant compression rates while reducing computational costs.

## I. Introduction

In the revolutionary classical information theory, Shannon solely focused on the statistics of the messages ignoring the semantics and the syntax to lay the foundations for the information age [1]. From Shannon's point of view, the fundamental communication problem is replicating messages from one point to another. This perspective completely ignores the intended goal of the replicated messages. However, with the advancement of communication networks, the quality of service (QoS) requirements, such as reliability, latency, security, and robustness, are getting more strict. We are reaching the capacity of our communication systems, and therefore, we must consider approaches that go beyond the traditional goal-agnostic message replication paradigm. This will require addressing the aforementioned strict QoS policies of massive and heterogeneous future communication systems. These approaches are beneficial for the low-latency and ultra-reliable communication systems for steering and control applications of the, e.g., Industry 4.0. This type of study can be referred to as goal-oriented or post-Shannon communication [2], [3]. The main idea of post-Shannon communication is to extend Shannon's communication framework to consider the semantics of the messages, the intended goals of communication, or both to reduce the required bandwidth of communication further. It is primarily applicable to the scenarios where the receivers have to achieve a particular goal or perform a specific task. Post-Shannon theory includes different approaches to address different types, such as message identification [4], [5], common randomness, functional compression (FC) [6], and medium as the message [7] for goal-oriented use cases.

In this paper, we solely focus on FC as a potential post-Shannon technique to reduce the required data rates for in-network computing applications for steering and control. The fundamental problem of FC is how much information the sources should transmit if the destination is interested in computing a function of the sources. FC can be seen as a generalisation of Shannon's source compression problem. In the traditional Shannon scenario, the primary purpose of data compression is to remove the redundancy of the source to reconstruct it as accurately as possible at the receiver end. In a sense, it answers the question of how much information the sources should transmit to compute the identity function (a function whose output is the same as its inputs) at the receiver. For example, for the distributed case of source compression, the Slepian-Wolf theorem describes the achievable compression rates [8]. In FC, however, the problem is generalised to reconstruct the output of any arbitrary function at the destination instead of solely focusing on regenerating the source information [9]. Therefore, the Slepian-Wolf source encoding example is a particular case of FC where the function to be computed at the receiver end is the identity function of the sources.

In this paper, we show, as a proof-of-concept, an implementation of FC for in-network control applications based on proportional–integral–derivative (PID) controllers. In our scenario, a PID controller located in the network is interested in computing a control function based on the sensor data of a plant. We use FC to compress the source's sensor data to transmit to the controller to calculate the control function, thus saving network resources.

The remaining of this paper is structured as follows. In Section II, we introduce the fundamentals of FC. In Section III, we describe our scenario of an in-network PID controller controlling the water level in a water tank. In Section IV we make a comparative analysis of two FC approaches for solving the problem. In Section V we discuss some open issues of FC for practical implementations, and finally, in Section VI we present our conclusions.

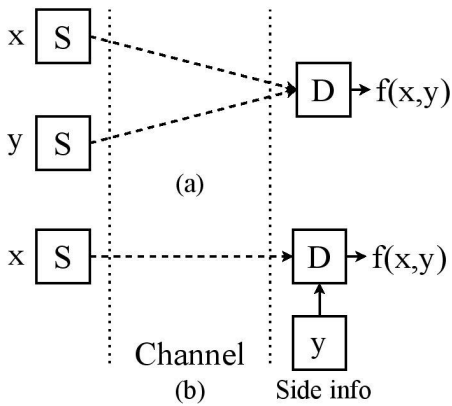

Fig. 1: FC scenario with two inputs.

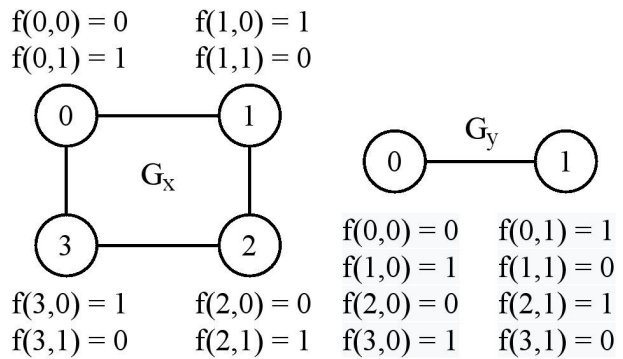

Fig. 2: Characteristic graphs for $x$ and $y$.

## II. FUNDAMENTALS OF FUNCTIONAL COMPRESSION

In this section, we highlight the fundamental principles of FC and its most commonly used coding technique based on graph colouring algorithms.

Let us assume a system consisting of two sources, $x$ and $y$, and a destination that wants to compute a function of the sources $f(x, y)$ as shown in Fig. 1. In the traditional Shannon scenario, the sources encode and transmit their messages to the destination via the wireless channel. The decoder at the destination decodes and replicates the values to compute the function $f(x, y)$ in Fig. 1(a). In another case, one of the sources can also be available as side information at the destination, and only one source transmits its message over the channel, as shown in Fig. 1(b). In traditional communication schemes, the sources would transmit the value of the variables, and the receiver would reconstruct them and compute the function of interest. In contrast, in an FC scenario, the sources encode and compress their message depending on the function the destination wants to compute, so the destination can obtain the output of its function upon receiving the compressed message. The most crucial factor of the FC is the encoding and compression happening at the source. The most common and practically feasible solution to address it is characterization graph colouring proposed in [9]. It is a method of forming a characteristic graph depending on all possible outcomes and labelling the graph with different colours for the same characteristics introduced by Körner in [10]. Körner also calculated the graph entropy of the characteristic graph based on the colour labels or chromatic number, which represents the minimum achievable information rate of communication symbol over a channel [10].

As an example, let us consider $x$ and $y$ are independent and uniformly distributed variables over $\{0, 1, 2, 3\}$ and $\{0, 1\}$, respectively, and the destination wants to compute $f(x, y) = (x + y) \bmod 2$. The sources can generate their characteristic graph $G_x$ and $G_y$ (as described in [6]) given that they know the function that the destination wants to compute beforehand. Fig. 2 shows the graphs $G_x$ and $G_y$ for all the possible outcomes of $f(x, y)$ for a given value of $x$ and all values of $y$ and for a given value of $y$ and all values of $x$, respectively. After calculating the outcomes, the edges between two vertices having different outcomes are drawn. The next step is to colour the graph, ensuring that the colours of the vertices having an edge cannot be the same. Fig. 3 shows the coloured graph for $x$ and $y$ where the chromatic number is 2 for both graphs. We can calculate the graph entropy from these graphs for $x$ and $y$ using (1) [10]. Here, $n$ is the power of the graph, and $H^x$ is the chromatic entropy of the colours based on the probability $p(c)$ of occurrence (2). For this example, the graph entropy for source $x$ is $H_G(x) = 1$, which means it has to send only 1 bit instead of 2 bits of information to the destination. Here, the compression rate is 50%. For source $y$, the graph entropy is $H_G(y) = 1$. No compression happens here because the minimum number of transmitting bits cannot be less than 1. After performing colour-coding, the sources send only the colour information to the destination decoder, which looks into a decoding lookup table to obtain the output of the function, as shown in Fig. 3. We can achieve more compression if we want to compute the function multiple times in batches. In this case, the power $n$ of the graph is increased. We can increase the power $n$ of a graph by performing the strong product of that graph with itself [6]. However, this multiplication method is unsuitable for low latency control applications where the function must be computed periodically. Thus, the case of $n > 1$ is excluded from this study.

$$\lim_{x \to \infty} \left( \frac{1}{n} H_{G^n}^x(\mathbf{x}) \right) = H_G(x) \qquad (1)$$

$$H^x = \sum p(c) \log_2 \left( \frac{1}{p(c)} \right) \qquad (2)$$

## III. DESCRIPTION OF OUR SCENARIO

In this section, we highlight different implementation scenarios for network FC (nFC) to evaluate FC from a network point of view. Let us assume a water tank with a water valve and a height measuring sensor. The water is leaking from a hole located at the bottom of the tank, as shown in Fig. 4. We want to control the water level in the tank to maintain a certain level. A PID controller is used to control the valve to release water into the tank to achieve the desired level. We consider the water level $h(t)$, and the desired level $s(t)$ is a time-based function as the water is constantly leaking. The PID controller requires three inputs to control the valve, i.e, proportional error: $e(t) = h(t) - s(t)$, integral error: $e_i(t) = \int_0^t e(\tau)d\tau$, and derivative error: $e_d(t) = \frac{de(t)}{dt}$. The PID controller collects this information from the sensor at each time $t$ wirelessly and computes the control function

$$f(e(t), e_i(t), e_d(t)) = k_p e(t) + k_i e_i(t) + k_d e_d(t), \quad (3)$$

to open the valve from $0\% - 100\%$ to maintain the desired water level. $k_p, k_i, k_d$ are the predefined tuning constants.

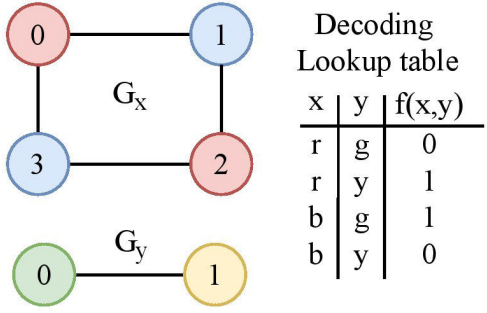

Fig. 3: Graph colouring and decoding lookup table.

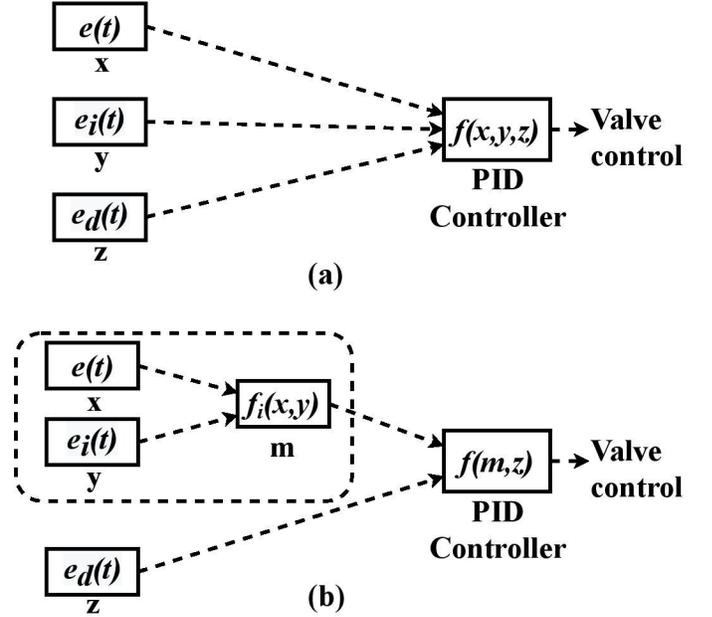

Fig. 5: (a) Simple nFC, and (b) cascaded nFC for water-tank scenario.

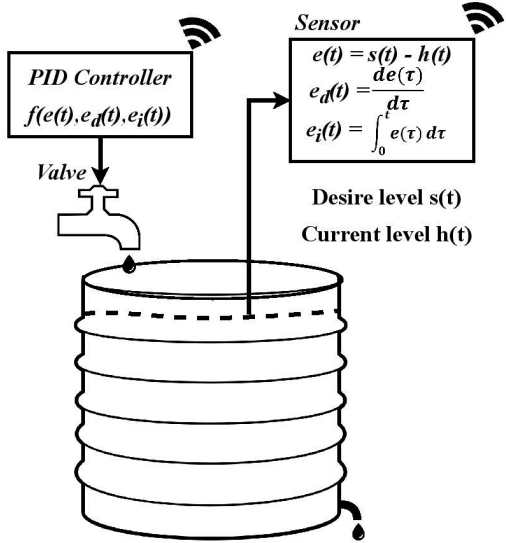

Fig. 4: Water tank leaking scenario.

Here, we can utilise FC in different ways to compute the PID controller function $f(e(t), e_i(t), e_d(t))$ for controlling the valve. Let us consider, $e(t)$, $e_i(t)$, and $e_d(t)$ are three different sources $x$, $y$, and $z$, respectively. These sources are uniformly and independently distributed over the sets of $n$-real numbers $\{x_1, x_2, \ldots, x_n\}$, $\{y_1, y_2, \ldots, y_n\}$, and $\{z_1, z_2, \ldots, z_n\}$, respectively. We can implement FC in two different ways to achieve the desired goal.

*a) Simple nF:* : The most straightforward and less complex way of implementing nFC is to generate a unique characteristic graph at each source to compress its information using graph colouring. After calculating the graph entropy, each source can transmit the colour information with the minimum number of bits to the PID controller as shown in Fig. 5(a). After receiving all the colour information from three different sources, the controller can decide how many valves should be opened from the lookup table to maintain the desired water level. However, this method has some drawbacks, such as high latency, a large lookup table, and high computational power. We will briefly discuss these drawbacks in section IV.

*b) Cascaded nFC:* : We can also use a cascaded form of nFC considering some intermediate nodes within the network. The main idea is to perform two-source FC multiple times until reaching the final destination. In this water tank scenario, we can consider two sources $x$ and $y$ perform FC and compute an intermediate function $f_i(x, y)$ at node $m$. Node $m$ has a range of values for $f_i(x, y)$ and becomes a new source. This can be easily done since the final target function is a linear function of the inputs. After that, $m$ and $z$ perform FC and compute the final function $f(m, z) = f(f_i(x, y), z) = f(x, y, z)$ as shown in Fig. 5(b). In this method, the output of the PID controller will be the same as in the previous method. However, this method requires less computational power and has a comparatively small lookup table and low latency. We will briefly discuss it in Section IV.

## IV. COMPARATIVE ANALYSIS

This section highlights the comparative analysis of the two different methods of nFC and addresses the open research issues from a network point of view. The parameters given in Table I are used to compare simple FC with cascaded FC. An Intel Core i7 11th generation 3.00GHz processor with 16GB memory is used to perform the whole simulation. We also consider noiseless channels among the sources and the destination.

TABLE I: Simulation parameters

| Parameter | Value |
| --- | --- |
| Tuning parameters, $k_p, K_i, K_d$ | $-0.5, 10, 90$ |
| Time steps, $T$ | 80 |
| Desired Level, $s(t)$ | $10m$ |
| Valve coefficient, $c_1$ | $50\ kg/s/\%$ open |
| Leaking coefficient, $c_2$ | $1\ kg/s$ open |
| Water density, $\rho$ | $1000\ kg/m^3$ |
| Graph colouring method | Greedy |

We model the water tank scenario using (4), where $c_1$ is the valve coefficient that represents the amount of water released per second per % of the open valve $v(t)$, and $c_2$ is the leaking coefficient that depends on the Bernoulli law for orifices [11]. Fig. 6 shows the simulation results, i.e., water level, valve opening, and error difference for the water tank scenario, considering the information size of each source is 7 bits. It is evident from the results that the PID controller can maintain the desired water level with both of the FC methods. However, minor distortions can be seen in the water level for the cascaded FC method owing to the cascading effect and quantization errors.

$$\frac{dh}{dt} = c_1 v(t) - c_2 \sqrt{h(t)} \quad (4)$$

We also simulate this water tank scenario for different information sizes of each source from 4 bits to 8 bits, as shown in Fig. 7 and Fig. 8. Fig. 7 shows the offline and online computation time comparison between simple FC and cascaded FC for different source sizes. Offline computational time represents the time taken to create encoding and decoding lookup tables with a greedy graph colouring algorithm for FC at the source and the destination, respectively. The online time refers to the time taken to access the lookup tables for FC during system operation. From the Fig. 7, it is clear that the cascaded FC requires much less offline and online computational time than the simple FC. The main reason is that the lookup tables are pretty small in cascaded FC due to cascading effect. In contrast, each node has to deal with huge lookup tables to keep track of all possible combinations of outcomes in simple FC.

$$Compression\ rate = 100 - \frac{Graph\ entropy \times 100}{Source\ size}\ \% \quad (5)$$

We also calculate and compare the compression rate of simple and cascaded FC in Fig. 8 for different information sizes of the sources. The compression rate represents the percentage of compression achieved by each method for a given information size of the sources. This rate depends on the graph entropy; the lower the entropy is, the higher the increment will be (5). From Fig. 8, we can see that the cascaded FC outperforms the simple FC. The simple FC also shows a downward trend in compression rate with a larger source's information size. Finally, it is noticeable from the results that cascaded FC is more suitable for low latency control applications than simple FC.

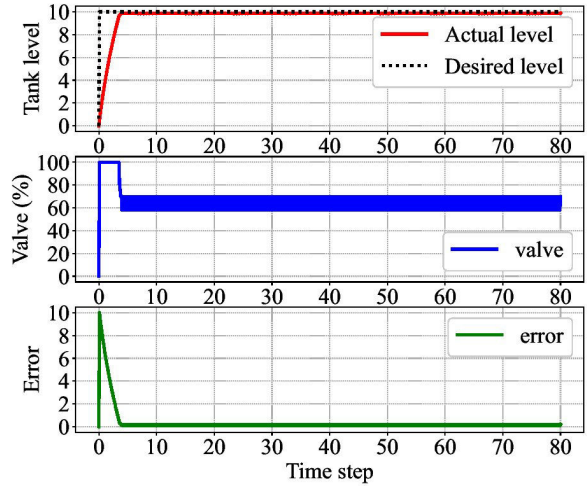

(a)

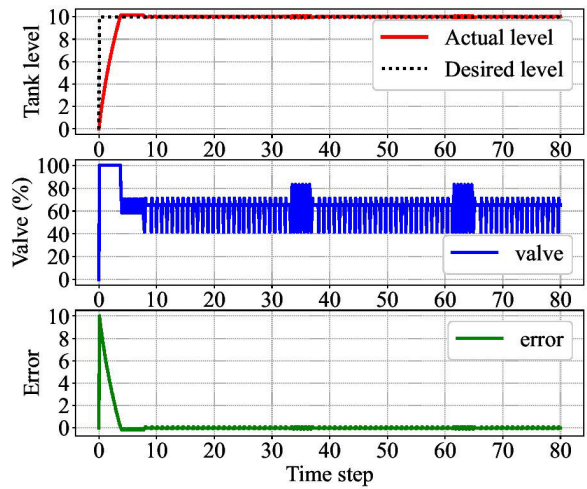

(b)

Fig. 6: Simulation results for the water tank scenario with (a) simple FC and (b) cascaded FC for 7 bits of each source.

V. OPEN ISSUES FOR PRACTICAL IMPLEMENTATIONS

This section discusses and emphasizes the unsolved problems from a practical point of view. nFC is a revolutionary goal-oriented communication technique that significantly impacts wireless networks with growing data rate requirements. However, there is plenty of space for improvements in nFC. One of the most crucial issues is the computational power of each network node, including the source and destination. The current approaches for code construction based on graph colouring have a high computational complexity that quickly grows with the size of the sources. Since the problem grows so fast, practical implementations like those described in this paper have to deal with distortions due to quantization errors. As shown in Fig. 6(b), we observed some distortion in the control output with a higher compression rate. There are some promising solutions. For instance, Malak proposed fractional graph colouring for FC algorithm with side information in [12], which is a less complex technique.

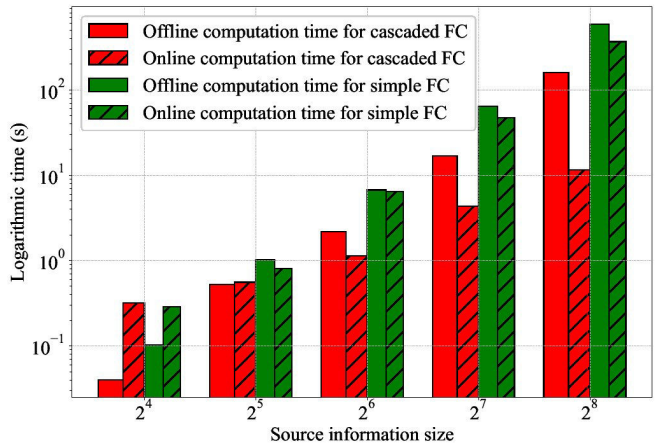

Fig. 7: Offline and online computational time comparison between simple FC and cascaded FC.

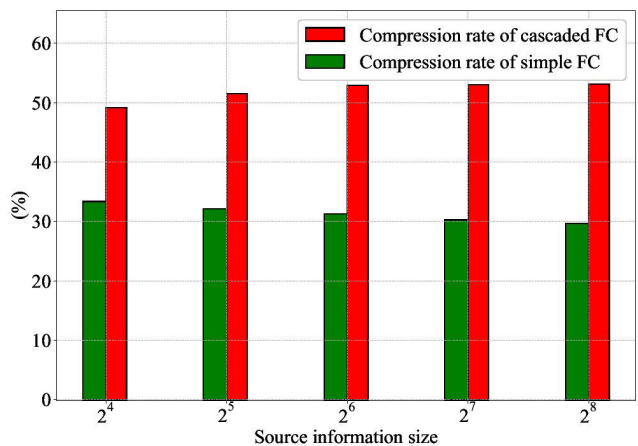

Fig. 8: Compression rate comparison between simple FC and cascaded FC.

## VI. CONCLUSION

In this paper, we presented a proof-of-concept implementation of FC as a technique to drastically reduce the data rates of goal-oriented communication as those employed by in-network control applications. We proposed and compared two approaches for network FC. We showed that for an in-network PID controller, it is possible to reduce the required data rates by over $50\%$. We showed how the size of the source information has a significant impact on the computational complexity of current FC approaches. We further discussed some of the open issues of FC related to practical implementations, especially those associated with the complexity of the algorithms.

## VII. ACKNOWLEDGEMENT

The authors acknowledge the financial support by the Federal Ministry of Education and Research of Germany in the programme of "Souverän. Digital. Vernetzt.". Joint project 6G-life, project identification number: 16KISK001K, and the German Research Foundation (DFG, Deutsche Forschungsgemeinschaft) as part of Germany's Excellence Strategy – EXC 2050/1 – Project ID 390696704 – Cluster of Excellence "Centre for Tactile Internet with Human-in-the-Loop" (CeTI) of Technische Universität Dresden.

## REFERENCES

[1] C. E. Shannon, "A mathematical theory of communication," *ACM SIGMOBILE mobile computing and communications review*, vol. 5, no. 1, pp. 3–55, 2001.
[2] J. A. Cabrera, H. Boche, C. Deppe, R. F. Schaefer, C. Scheunert, and F. H. Fitzek, "6g and the post-shannon theory," *Shaping Future 6G Networks: Needs, Impacts, and Technologies*, pp. 271–294, 2021.
[3] R. Bassoli, F. Fitzek, and E. C. Strinati, "Why do we need 6g?" *ITU Journal on Future and Evolving Technologies*, vol. 2, no. 6, pp. 1–31, 2021.
[4] C. v. Lengerke, A. Hefele, J. A. Cabrera, and F. H. P. Fitzek, "Stopping the data flood: Post-shannon traffic reduction in digital-twins applications," in *NOMS 2022-2022 IEEE/IFIP Network Operations and Management Symposium*, 2022, pp. 1–5.
[5] R. Ahlswede and G. Dueck, "Identification via channels," *IEEE Transactions on Information Theory*, vol. 35, no. 1, pp. 15–29, 1989.
[6] S. Feizi and M. Médard, "On network functional compression," *IEEE transactions on information theory*, vol. 60, no. 9, pp. 5387–5401, 2014.
[7] F. H. Fitzek, "The medium is the message," in *2006 IEEE International Conference on Communications*, vol. 11. IEEE, 2006, pp. 5016–5021.
[8] D. Slepian and J. Wolf, "Noiseless coding of correlated information sources," *IEEE Transactions on Information Theory*, vol. 19, no. 4, pp. 471–480, 1973.
[9] V. Doshi, D. Shah, M. Médard, and M. Effros, "Functional compression through graph coloring," *IEEE Transactions on Information Theory*, vol. 56, no. 8, pp. 3901–3917, 2010.
[10] J. Körner, "Coding of an information source having ambiguous alphabet and the entropy of graphs," in *6th Prague conference on information theory*, 1973, pp. 411–425.
[11] G. Mikhailov, "Daniel bernoulli, hydrodynamica (1738)," in *Landmark Writings in Western Mathematics 1640-1940*. Elsevier, 2005, pp. 131–142.
[12] D. Malak, "Fractional graph coloring for functional compression with side information," *arXiv preprint arXiv:2204.11927*, 2022.